\documentclass[12pt]{article}
\usepackage{epsfig,amsfonts,amssymb}
\usepackage{amsmath,graphics}
\usepackage[parfill]{parskip}
\usepackage{color}

\usepackage{hyperref}
\def\ZZZ{{\hbox{ Z\kern-1.6mm Z}}}
\def\RRR{{\hbox{ R\kern-2.4mm R}}}
\def\CCC{{\hbox{ C\kern-2.0mm C}}}
\def\zzz{{\hbox{z\kern-1mm z}}}
\newcommand{\mathsym}[1]{{}}

\newcommand{\qeq}{{\hbox{=\kern-2.3mm ? \kern.5mm }}}
\renewcommand{\qeq}{=}

\newcommand{\be}{\begin{equation}}
\newcommand{\ee}{\end{equation}}
\newcommand{\ben}{\begin{eqnarray}\displaystyle}
\newcommand{\een}{\end{eqnarray}}

\newcommand{\sectiono}[1]{\section{#1}\setcounter{equation}{0}}

\def\one{{\hbox{ 1\kern-.8mm l}}}
\def\zero{{\hbox{ 0\kern-1.5mm 0}}}

\begin{document}
\baselineskip 24pt

\begin{center}
{\Large \bf Penrose Limit and Non-relativistic geometries
}

\end{center}

\vskip .6cm
\medskip

\vspace*{4.0ex}

\baselineskip=18pt

\centerline{\large \rm   Abhishek Mathur and Yogesh K. Srivastava}

\vspace*{4.0ex}

\centerline{\large \it National Institute of Science Education and Research}

\centerline{\large \it  Sachivalaya Marg,
Bhubaneswar 751005, INDIA}

\vspace*{1.0ex}
\centerline{E-mail: abhishekmathur, yogeshs@niser.ac.in }

\vspace*{5.0ex}

\centerline{\bf Abstract} \bigskip

For the AdS/CFT duality, considerations of plane wave metric which is obtained as Penrose limit of $AdS_5 \times S^5$ proved to be
quite useful and interesting. In this work, we obtain Penrose limit metrics for Lifshitz, Schrodinger, hyperscaling violating Lifshitz and hyperscaling violating Schrodinger geometries. These geometries usually contain singularities for certain range of parameters and we discuss how these singularities appear in the Penrose limit metric.  For some cases, there are non-singular metrics possible for certain parameter values and the metric can be extended beyond the coordinate singularity, as discussed in many previous works. Corresponding Penrose limit metrics also display similar features. For the hyperscaling violating Schrodinger metric, we obtain metric extension for some cases.
\vfill \eject

\baselineskip=18pt

\tableofcontents

\section{Introduction} \label{s1}

Holographic duality has been one of the most fruitful recent ideas in theoretical physics. Dual gravitational description of certain 
strongly coupled field theories gives a new tool to tackle otherwise difficult strong coupling regime.  Strong coupling questions can sometimes be translated to weakly coupled gravitational problems in bulk. In the last few years, there have been rapid advances in 
applying it to cases far removed from the original (and still the best understood) example of $AdS_5$/$CFT_4$. In particular, after the pioneering work
of \cite{Son:2008ye, Balasubramanian:2008dm, Kachru:2008yh} there has been a lot of work on gravitational duals of non-relativistic systems.
Using gauge/gravity duality one expects to be able to probe strong coupling limit of these field theories.
Lifshitz and Schrodinger geometries represent gravitational duals of systems with anisotropic scaling $(t,x^i) \rightarrow (\lambda^z t, \lambda x^i)$.
In the case of $AdS$/$CFT$, it proved fruitful to consider Penrose limit of both sides of the correspondence. Since string theory on pp-wave spacetimes
can be quantized, this allowed for a more thorough check of duality \cite{Berenstein:2002jq}. Based on these considerations, we obtain Penrose limit 
of Lifshitz and Schrodinger type geometries. The field theory duals of these systems are not that well understood so we will have much less to say about the other side of duality. If non-relativistic duality is correct then string theory on plane wave metrics that we obtain should be dual to some sector of states on field theory side. Lifshitz metrics are known to have pp-curvature singularities (infinite tidal force for freely falling observers) \cite{Kachru:2008yh,Copsey:2010ya,Horowitz:2011gh}. Schrodinger geometries do not suffer from such singularities (for $z \geq 2$) and for such cases, smooth extensions were constructed in \cite{Blau:2009gd}. In the plane wave limits that we find, we will see these singularities in the Lifshitz case and their absence in the Schrodinger case. In addition to Lifshitz and Schrodinger spacetimes, we also consider Penrose limits of hyperscaling violating geometries considered in \cite{Ogawa:2011bz,Dong:2012se}. Among the class of hyperscaling violating geometries, some of them have smooth extensions while others have pp-curvature singularities \cite{Lei:2013apa}. Penrose  limit of these geometries also exhibits these features. In plane wave metrics obtained after taking Penrose limit, all the information is contained in profile function and singularities of the metric occur where 
profile function is singular.

Penrose limit has been discussed extensively since the original work of Penrose \cite{Penrose}. In \cite{Blau:2002mw} and many subsequent papers, \cite{Blau:2003dz,Blau:2004yi}, the process of obtaining Penrose limit has been spelt out in great detail. We will use the method given in these works to obtain the Penrose limit.

Even though geometries like Lifshitz and Schrodinger give dual description of non-relativistic field theories and the Penrose limit is associated with 
observers in a highly boosted frame, we find that the limiting metrics we get are not pathological. Hence, as  easily tractable  limits, plane wave spacetimes that we obtain are of particular interest.

This paper is organized as follows. In the next section, we review some of the metrics dual to non-relativistic systems which have been discussed in 
recent years. In section $3$, we obtain Penrose limit of the Lifshitz metric. We derive the Penrose limit metric by two different methods. Traditional one is 
based on first converting the metric to Penrose adapted coordinates and then taking Penrose limit. The metric one gets is in Rosen coordinates which may contain spurious coordinate singularities. It is usually better to work with Brinkmann coordinates in which plane wave metric is characterised by a single symmetric matrix-valued function $A_{ab}(u)$. We perform the coordinate transformations to get the Penrose limit Lifshitz metric in Brinkmann form. In
 \cite{Blau:2003dz}), a covariant characterisation of Penrose limit was given which directly gives us the Penrose limit metric in Brinkmann coordinates 
and hence wave profile $A_{ab}(u)$. In section $3.2$, we derive Penrose limit metric for Lifshitz case using this covariant method and see that wave 
profile matches with what we got by the traditional method. In section $3.3$, we note that Penrose limit of Lifshitz metric gives plane wave metric 
which is identical to that obtained in \cite{Horowitz:2011gh} by Horowitz and Way after taking the near singularity limit of the Lifshitz metric (in their paper, it's not presented as Penrose limit metric). In section $4$, we find Penrose limit of Schrodinger metric and from the profile function $A_{ab}$, it's clear that $z<2$ case is singular while $z \geq 2$ case is regular. In section $5$, we find the Penrose limit of hyperscaling violating 
Lifshitz metrics. Here there are two parameters, dyanmical exponent  and hyperscaling violating parameter and hence the structure of Penrose limit 
metric is more interesting. We can consider various limits and identify ranges where singularities occur. In section $6$, we consider hyperscaling violating Schrodinger spacetimes. As in previous case, there are two parameters and hence one has to consider various ranges for which singularities occur. In section $6.1$, we derive the constraints from null energy condition on dynamical exponent $z$ and hyperscaling violating parameter $k$. In section $6.2$, we derive Penrose limit for this case and consider possible singularities. For non-singular cases, metric extensions for Schrodinger and hyperscaling violating Lifshitz metrics have been constructed in \cite{Blau:2009gd} and \cite{Lei:2013apa} respectively. In section $6.3$, we construct
 smooth coordinate system for the case when $z$ and $k$ both are equal to $2$. In section $7$, we conclude and discuss future directions. In appendix A,
 we review some results about Penrose limits, drawing heavily on \cite{Blaureview}. In appendix B, we consider Penrose limit of hyperscaling
 violating Lifshitz metric in a slightly different coordinate system. Unlike section $5$, we use covariant method for Penrose limit here. 
\section{ Non-relativistic field theories and their dual metrics} \label{s2}

Many non-relativistic conformal field theories (CFT) have been found useful in condensed matter physics (e.g, \cite{Nishida:2007pj}) and many other areas of physics. Examples of such systems would be fermions at unitarity and theories at Lifshitz point. In general these systems have non-trivial scaling properties rather than just scale invariance. Dual gravitational descriptions of these have been explored actively in past few years and simplest of these are Lifshitz and Schrodinger metrics

\begin{eqnarray}
ds^2 |_{Lif} =  \left( -r^{2z}dt^2 + \frac{dr^2}{r^2} + r^2 dx_i dx^i \right) \\
ds^2 |_{Sch} =-r^{2z}dt^2+r^2\left(-2dtd\zeta+dx_i dx^i \right)+\frac{dr^2}{r^2}
\end{eqnarray}

Here $z$ is called the dynamical critical exponent and $i=1,...d$. Case $z=1$ corresponds to AdS metric in Poincare coordinates with $AdS$ radius unity.  For condensed matter applications, in addition to these, hyperscaling violating Lifshitz and 
Schrodinger metrics have also been considered. Without hyperscaling violation, the entropy of dual field theory scales as $S \sim T^{d/z}$ where $T$ is temperature and $d$ is the number of spatial dimensions. Due to random field fluctuations, hyperscaling is violated in many condensed matter systems. Gravitational duals of these reflect this altered scaling.  For the Lifshitz case
\begin{equation}
ds^2=\frac{1}{r^{2\theta/d}}\left(-r^{2z}dt^2+\frac{dr^2}{r^2}+r^2dx_i dx^i \right)
\end{equation}

Non-zero $\theta$ parametrizes the deviation from naive scaling for entropy in field theory. Similarly, for the Schrodinger case, we have 
\begin{equation}
ds^2=r^{-k}\left\{-r^{2z}dt^2+r^2\left(-2dtd\zeta+dx_i dx^i \right)+\frac{dr^2}{r^2}\right\}
\end{equation}

Here $k$ plays the same role as $2\theta/d$ in Lifshitz case above.
These metrics are not scale invariant but transform with a constant conformal factor under scale transformation. For example, in the Lifshitz case, under a scaling $(t,x^i,r) \rightarrow (\lambda^z t, \lambda x^i , r/\lambda)$, the metric scales as $ds^2 \rightarrow \lambda^{\frac{2\theta}{d}}ds^2$.
Roughly speaking, such theories with hyperscaling violation have thermodynamic behavior of a theory with scaling $z$ but living in $d-\theta$ dimensions as discussed in \cite{Dong:2012se}.

\section{Penrose Limit of Lifshitz metric} \label{s3}
\setcounter{equation}{0}
In this section, we  construct the Penrose limit of the Lifshitz metric. As is known \cite{Kachru:2008yh,Horowitz:2011gh}, Lifshitz metric has a null singularity at $r=0$. We will see that the corresponding Penrose limit reflects this. As an illustration, we will calculate Penrose limit using two different methods: one using Penrose adapted coordinates and second using a covariant method as presented in \cite{Blau:2003dz}. First method gives the plane wave metric in Rosen coordinates which we convert to Brinkmann form while the covariant method gives the metric directly in Brinkmann form. For completeness, some details about construction of the Penrose limit metric are given in appendix A.

\subsection{Penrose limit via adapted coordinates} \label{s3.1}

We start with Lifshitz metric
\be
ds^2 =   -r^{2z}dt^2 + \frac{dr^2}{r^2} + r^2  dx_i dx_i 
\ee

In this method for obtaining Penrose limit, as reviewed in appendix A, we first transform the metric to Penrose adapted coordinates which are coordinates adapted to chosen null geodesic, say $\gamma$
\begin{equation}
ds_\gamma^2=2dUdV+a(U,V,Y^k)dV^2+2b_i(U,V,Y^k)dVdY^k+g_{ij}(U,V,Y^k)dY^idY^j
\end{equation}

Null geodesics are given by
\be
-r^{2z} \dot{t}^{2} + \frac{\dot{r}^{2}}{r^2} + r^2 \dot{x}_{i}\dot{x}_{i} =0
\ee

Here derivativeis with respect to affine parameter $\tau$. This Lagrangian gives $\dot{t}= -\frac{E}{r^{2z}}$ and  $\dot{x}_{i} = \frac{p_i}{ r^2}$. Imposing nullity yields
\be
-\frac{E^{2}}{r^{2z}} + \frac{\dot{r}^{2}}{r^2} + \frac{p^{2}}{r^2} =0
\ee 

where  $p^2 = \sum_{i}p_i p_i$. Radial coordinate thus satisfies
\be
\left(\frac{dr}{d\tau}\right)^2 = r^2 \left(\frac{E^{2}}{r^{2z}} -\frac{p^{2}}{r^2}\right)
\ee

For simplicity, we take only one of $p_i$, say $p_1$ to  be non-zero.  To go to adapted coordinates, we choose $\tau$ as new coordinate $U$. As discussed in \cite{Blaureview}, we take the solution of Hamilton-Jacobi(HJ) equation
$g^{\mu\nu}\partial_{\mu}S\partial_{\nu}S =0$ as another null coordinate $V$ i.e. $S=V$. Based on isometries, we substitute the ansatz 
$S= -Et + p_1 x_1 + \rho(r)$. Putting this in HJ equation, we get
\be
g^{tt}\left(\frac{\partial S}{\partial t}\right)^2 + g^{ij}\left(\frac{\partial S}{\partial x_i}\right) \left(\frac{\partial S}{\partial x_j}\right)+ g^{rr}\rho'^2 =0
\ee

This gives, using null condition, 
\be
\rho' =\frac{d\rho}{dr}= \frac{1}{r^2}\frac{dr}{d\tau}  \ \ ,  \ \ \frac{d\rho}{d\tau}= \frac{1}{r^2}\left(\frac{dr}{d\tau}\right)^2
\ee

 Since $x_1 = X + \int \frac{p_{1}dU}{r^2}$, we take integration constant $X$ as new coordinate. Using 
\be
dr = \dot{r}(U) dU, \ \ \ \ \  \ \ \ dx_1 = dX + \frac{p_{1}dU}{r^2}
\ee
\be
dV =dS = -Edt + p_{1}\left(dX+ \frac{p_{1}dU}{r^2}\right) + dU\left(\frac{E^{2}}{r^{2z}} -\frac{p_{1}^{2}}{r^2}\right)
\ee

Putting these in the Lifshitz metric, we get
\be
ds^2 =  2dUdV + 2\frac{p_1}{E^2}r^{2z}dVdX  -\frac{r^{2z}}{E^2}dV^2 + r^2 \sum_{i=2}^n d\tilde{x}_{i}^{2} + dX^2(r^2 -r^{2z}\frac{p_{1}^2}{E^2} )
\ee

Here $r=r(U)$. Taking the Penrose limit, as described in the appendix,  we get
\begin{eqnarray}
ds^2&=&2dUdV +\left[\left(r^2-\frac{p_{1}^2}{E^2}r^{2z}\right)dX^2+r^2\sum_{i=2}^ndx_i^2\right]\nonumber \\
&=&2dUdV+\left(\frac{\dot{r}r^z}{E}\right)^2dX^2+ r^2\sum_{i=2}^ndx_i^2
\end{eqnarray}

Converting from Rosen to Brinkmann coordinates, we have
\begin{eqnarray}
A_{11}=\frac{\ddot{e_1}}{e_1}&=&\frac{1}{\dot{r}r^z}\frac{d^2}{dU^2}(\dot{r}r^z) = \frac{z(1-z)p_{1}^2}{r^2}+\frac{(1-z)E^2}{r^{2z}}\\
A_{ii}=\frac{\ddot{e_i}}{e_i}&=&\frac{\ddot{r}}{r} = \frac{(1-z)E^2}{r^{2z}}
\end{eqnarray}

We see that for $AdS$ (i.e. $z=1$)  the Penrose limit metric is flat space as expected.  In the next section we will derive these equations directly using the covariant method.  

\subsection{Penrose limit using covariant method} \label{s3.2}

In this section, we use the covariant characterization of Penrose limit as given in \cite{Blau:2003dz}. In this approach, we directly arrive at the Penrose limit metric in Brinkmann coordinates and the plane wave is described by a wave profile $A_{ij}$. Again, we start with Lifshitz metric
\begin{equation}
ds^2=\left(-r^{2z}dt^2+\frac{dr^2}{r^2}+r^2d\vec{x}^2\right)
\end{equation}
where $\vec{x}$ represents a vector in the $n$ dimensional flat subspace ($d\vec{x}^2=dx_1^2+\dots+dx_n^2$. So the total spacetime dimension is $D=n+2$. Geodesic Lagrangian in this spacetime is given by
\begin{equation}
{\cal L}=\frac{1}{2}\left(-r^{2z}\dot{t}^2+\frac{\dot{r}^2}{r^2}+r^2\dot{\vec{x}}^2\right)
\end{equation}
Here, derivative is with resepct to affine parameter $U$. We can choose the transverse coordinates $(x_1,x_2,\dots,x_n)$ such that $\dot{\vec{x}}=\dot{x_1}$ and $\dot{x}_2=\dot{x}_3=\dots=\dot{x}_n=0$. The Lagrangian then becomes
\begin{equation}
{\cal L}=\frac{1}{2}\left(-r^{2z}\dot{t}^2+\frac{\dot{r}^2}{r^2}+r^2\dot{x_1}^2\right)
\end{equation}
The Euler Lagrange equations give
\begin{equation}
\dot{t}=\frac{E}{r^{2z}}
\end{equation}
where $E$ is a constant. For the $x_1$ coordinate, we get
\begin{equation}
\dot{x_1}=\frac{P}{r^2}
\end{equation}
where $P$ is a constant (same as $p_1$). Using the condition for the null geodesic $({\cal L}=0)$, we will get the equation for $r$.
\begin{equation}
\dot{r}^2= \left(\frac{E^2}{r^{2(z-1)}}-P^2\right) \label{rgeodesic}
\end{equation}
Now we need to find the Hamilton Jacobi function, which we calculated in previous section
\begin{equation}
S=-Et+Px_1+\rho(r)
\end{equation}

Using the earlier result
\begin{equation}
\rho(r)=\int\frac{1}{r^2}\dot{r}^2du= \int\left(\frac{E^2}{r^{2z}}-\frac{P^2}{r^2}\right)du
\end{equation}

where we now use $U=u$ since final Brinkmann coordinate will be $u$.
Next we need to write $ds^2$ in a parallel frame
\begin{equation}
ds^2=2E^+E^-+\delta_{ab}E^aE^b
\end{equation}

We construct the parallelly propagated frame as follows:
\begin{equation}
E_+=\dot{t}\partial_t+\dot{r}\partial_r+\dot{x_1}\partial_{x_1} \ \ , \ \ E_+ |_{\gamma}=\partial_u
\end{equation}
Since there is no evolution in the $x_a$ directions, for $a=2,3\dots,n$ we can have
\begin{equation}
E_{a}=rdx_a
\end{equation}

We can leave $E_{-}$ unspecified for present. It can be shown that all these terms satisfy $\nabla_uE^a_\mu=0$. To calculate the wave profile $A_{ab}$, 
we calculate $B_{ab}$ as reviewed in appendix A. 
\begin{equation}
B_{ab}=E_a^\mu E_b^\nu \nabla_\mu\partial_\nu S
\end{equation}
For our case, 
\begin{eqnarray}
B_{ab}&=&E_a^{x_a}E_b^{x_b}\nabla_{x_a}\partial_{x_b}S\\
&=&E_a^{x_a}E_b^{x_b}\left(\partial_{x_a}\partial_{x_b}S- \Gamma^\alpha_{x_ax_b}\partial_\alpha S\right)\\
&=&-E_a^{x_a}E_b^{x_b}\Gamma^\alpha_{x_ax_b}\partial_\alpha S
\end{eqnarray}
 Therefore, substituting the values, we get 
\begin{equation}
B_{ab}= \delta_{ab}\partial_u\log r
\end{equation}

We also have the condition (see appendix A),
\begin{equation}
Tr(B)=\frac{1}{\sqrt{-g}}\partial_\mu(\sqrt{-g}\dot{x}^\mu)
\end{equation}

Evaluating this, we get
\begin{equation}
Tr(B)=\partial_u\log(\dot{r}r^{n+z-1})
\end{equation}
Hence
\begin{equation}
B_{11}=\partial_u\log(r^z\dot{r})
\end{equation}
It can easily be shown that $B_{ab}$ is diagonal. So in calculating profile function using 
\begin{equation}
A_{ab}=\frac{d}{du}B_{ab}+(B_{ab})^2
\end{equation}
we use $B_{ab}=\delta_{ab}\partial_u\log K_a$ (no sum over $a$) to get 

\begin{equation}
A_{ab}=\delta_{ab}K_a^{-1}\partial_u^2K_a
\end{equation}
where for $a=1$, $K_a=r^z\dot{r}$ and for $a=2\text{ to }n$, $K_a=r$.\\
Using the geodesic equation for $r$ we get
\begin{equation}
\ddot{r}=\frac{(1-z)E^2}{r^{2z-1}}
\end{equation}
and
\begin{equation}
\frac{\dddot{r}}{\dot{r}}=\frac{(1-z)(1-2z)E^2}{r^{2z}}
\end{equation}
\begin{equation}
A_{22}=A_{33}=\dots=A_{nn}=\frac{\ddot{r}}{r}=\frac{(1-z)E^2}{r^{2z}}
\end{equation}
\begin{eqnarray}
A_{11}&=&\frac{1}{\dot{r}r^z}\partial_u^2(\dot{r}r^z)\\
&=&\frac{1}{\dot{r}r^z}\partial_u\left[\ddot{r}r^z+zr^{z-1}\dot{r}^2\right]\\
&=&\frac{1}{\dot{r}r^z}\left[\dddot{r}r^z+3zr^{z-1}\dot{r}\ddot{r}+z(z-1)r^{z-2}\dot{r}\right]\\
&=&\frac{\dddot{r}}{\dot{r}}+3z\frac{\ddot{r}}{r}+z(z-1)\frac{\dot{r}^2}{r^2}
\end{eqnarray}
Putting the values of $\dddot{r}$ and $\ddot{r}$ as shown above we get
\begin{equation}
A_{11}=\frac{z(1-z)P^2}{r^2}+\frac{(1-z)E^2}{r^{2z}}
\end{equation}
The case $z=0$ gives the flat metric.  Wave profile diverges at $r(u)=0$ which indicates the presence of a true singularity at that point, as initially parallel null geodesic diverges at the point of divergence of the Penrose limit wave profile $A_{ab}$.\\ 
As $r \rightarrow 0$, we see that second term in $A_{ab}$ dominates (for $z \geq 1$ ) and also in the radial geodesic equation, we have
\begin{equation}
\dot{r}=\frac{E}{r^{z-1}}
\end{equation}
Solving the above equation for $r(u)$ we will get
\begin{equation}
u=\frac{r^z}{Ez}
\end{equation}

On substituting this in the expression of the wave profile $A_{ab}$, we get
\begin{equation}
A_{ab}=\frac{1-z}{z^2u^2}\delta_{ab}
\end{equation}
So in Brinkmann coordinates, the Penrose limit metric looks like
\begin{equation}
ds^2=2dudv+\frac{1-z}{z^2u^2}x_i^2du^2+d\vec{x}^2  \label{lif in br}
\end{equation}
where $x_i's$ are the transverse coordinates. This is consistent with the behaviour of the Penrose limit metric near the singularity as discussed in \cite{Blau:2004yi}. Dominant singularity is $u^{-2}$ even though there are subdominant pieces with weaker singularity $u^{-\alpha}$ with $\alpha <2$. Null energy condition for Lifshitz metric requires $z \geq 1$ and so we only need to consider this case. From the geodesic equation \ref{rgeodesic},
one can see that we can not have $E =0$ and $P \neq 0$. For the Lifshitz metric, null energy condition requires $z>1$ and so $A_{ab}$ are negative 
for cases satisfying the null energy condition.

\subsection{ Penrose limit and the near singularity limit of Lifshitz spacetime} \label{s3.3}
In \cite{Horowitz:2011gh}, Horowitz and Way take the near singularity limit of the Lifshitz metric
\begin{equation}
ds^2=l^2\left(-r^{2z}dt^2+\frac{dr^2}{r^2}+r^2d\vec{x}^2\right)
\end{equation}
 
and show that in the appropriate limit, they get a plane wave metric. They use this plane wave metric to show that string passing through such 
a singularity will be infinitely excited. The Penrose limit metric for the radial null geodesic that we have calculated, via different methods, turns out to be same as the near singularity limit calculated by Horowitz and Way. We quickly review their construction to show this. 
Penrose limit metric for the radial null geodesic was found to be
\begin{equation}
ds^2=2dudv+\frac{1-z}{z^2u^2}x_i^2du^2+d\vec{x}^2  
\end{equation}
To find the near singularity limit we define the tortoise coordinate $r_*$ such that
\begin{equation}
r_*=-\frac{1}{2r^z}\Rightarrow dr_*=r^{-1-z}dr
\end{equation}
and define null coordinates
\begin{equation}
U=t-r_*,\;\;\;V=t+r_*
\end{equation}
The metric becomes
\begin{equation}
ds^2=-\frac{4l^2}{z^2(U-V)^2}dUdV+l^2\left(\frac{4}{z^2(U-V)^2}\right)^{1/z}d\vec{x}^2
\end{equation}
Here in the coordinate transformation we have
\begin{equation}
r^z=-\frac{1}{zr_*}=\frac{2}{z(U-V)}
\end{equation}
For the near singularity limit i.e. $r\rightarrow0$ we have $U>>V$. So the metric in the above equation becomes
\begin{equation}
ds^2\approx-\frac{4l^2}{z^2U^2}dUdV+l^2\left(\frac{4}{z^2U^2}\right)^{1/z}d\vec{x}^2
\end{equation}
Now let $U=-4l^2/z^2u$ under which a small $r$ can be approximated by
\begin{equation}
r^z\approx-\frac{zu}{2l^2}
\end{equation}
So our line element becomes
\begin{equation}
ds^2\approx-dudV+l^2\left(\frac{zu}{2l^2}\right)^{2/z}d\vec{x}^2
\end{equation}
Defining
\begin{equation}
x_i=\frac{X_i}{l}\left(\frac{zu}{2l^2}\right)^{-1/z}\;\;and\;\;V=v-\frac{1}{zu}X_iX^i
\end{equation}
we get
\begin{equation}
ds^2\approx-dudv+\frac{1-z}{z^2u^2}X_iX^idu^2+d\vec{X}^2
\end{equation}
This is same as the Penrose limit of the Lifshitz metric (after a change in sign to match the conventions). So the near singularity limit of Horowitz 
 and Way is the same as the Penrose limit metric. String propagation in this near singularity limit was considered by Horowitz and Way and same results 
will hold in our case. Strings become infinitely excited and hence $r=0$ is a real singularity in string theory as well.

\section{Penrose limit of Schrodinger Metric} \label{s4}
\setcounter{equation}{0}

In this section, we derive plane wave metric corresponding to Schrodinger metric. Unlike the Lifshitz metric, Schrodinger metric is non-singular at the horizon $r= 0$ for $z \geq 2$ and metric extensions have been constructed for these cases \cite{Blau:2009gd}.  We will see that the Penrose limit metric shares this and makes it manifest. We start with $d+3$ dimensional Schrodinger metric

\begin{equation}
ds^2=-r^{2z}dt^2+r^2\left(-2dtd\zeta+d\vec{x}^2\right)+\frac{dr^2}{r^2}
\end{equation}
where $\vec{x}$ represents a vector in $d$ dimensional flat Euclidean space. 
In this case we will include $t$, $r$, $\zeta$ and one of the transverse coordinates $x_1$ in our geodesic. So the Lagrangian in this case will be
\begin{equation}
{\cal L}=\frac{1}{2}\left(-r^{2z}\dot{t}^2-2r^2\dot{t}\dot{\zeta}+\frac{\dot{r}^2}{r^2}+r^2\dot{x}_1^2\right)
\end{equation}
Here, derivative is with respect to affine parameter $U$. Solving the Euler Lagrange equation for $t$ we get
\begin{equation}
r^{2z}\dot{t}+r^2\dot{\zeta}=E
\end{equation}

Similarly, we get geodesic equations for $\zeta$ and $x_2$.
\begin{eqnarray}
\dot{t}&=&\frac{M}{r^2}\\
\dot{\zeta}&=&\frac{1}{r^2}\left(E-Mr^{2(z-1)}\right)\\
\dot{x}_1&=&\frac{P}{r^2}
\end{eqnarray}
We will use the null condition i.e. ${\cal L}=0$ instead of $r$ geodesic equation.
\begin{equation}
\dot{r}^{2}=2ME-P^2-M^2r^{2(z-1)} \label{dotrsm}
\end{equation}

Considering $r=r(U)$, we have
\begin{equation}
dr=\dot{r}dU
\end{equation}
For $t$ equation, we get 
\begin{equation}
t=\int\frac{M}{r^2}dU+T
\end{equation}
Similarly, for the other coordinates we have
\begin{eqnarray}
d\zeta&=&\frac{1}{r^2}\left(E-Mr^{2(z-1)}\right)dU+d\lambda\\
dx_1&=&\frac{P}{r^2}dU+dX_1
\end{eqnarray}
Substituting these expressions of $dt$, $d\zeta$, $dx_1$ and $dr$ in the metric we get
\begin{equation}
ds^2=-2EdUdT-2MdUd\lambda+2PdUdX_1-r^{2z}dT^2-2r^2dTd\lambda+r^2dX_1^2+r^2 dx_i^2
\end{equation}
where $i=2,...d$. Define
\begin{equation}
V=-ET-M\lambda+PX_1
\end{equation}
Replacing $\lambda$ with $V$, we will get
\begin{equation}
ds^2=2dUdV+\frac{2r^2}{M}dVdT+\left(\frac{2E}{M}r^2-r^{2z}\right)dT^2-\frac{2Pr^2}{M}dTdX_1+r^2dX_1^2+r^2dx_i^2
\end{equation}
Defining
\begin{equation}
X=X_1-\frac{P}{M}T
\end{equation}
and replacing it with $X_1$ in the metric above we will get
\begin{eqnarray}
ds^2&=&2dUdV+\left(\frac{2ME-P^2}{M^2}r^2-r^{2z}\right)dT^2+r^2dX^2+r^2 dx_i^2\\
ds^2&=&2dUdV+\frac{r^2\dot{r}^2}{M^2}dT^2+r^2dX^2+r^2 dx_i^2
\end{eqnarray}
Remembering that $r=r(U)$, we have the plane wave metric in Rosen coordinates. We now convert this to Brinkmann coordinates. Using vielbeins 
$e_0=\frac{r\dot{r}}{M}$ and $e_1=e_2=\dots=e_n=r$, for the diagonal metric,  we get
\begin{eqnarray}
A_{00}=\frac{\ddot{e_0}}{e_0}&=&\frac{\dddot{r}}{\dot{r}}+3\frac{\ddot{r}}{r}\\
A_{ii}=\frac{\ddot{e_i}}{e_i}&=&\frac{\ddot{r}}{r}
\end{eqnarray}
Using equation(\ref{dotrsm}) we get
\begin{equation}
\ddot{r}= -(z-1)r^{2z-3}M^2
\end{equation}
and
\begin{equation}
\frac{\dddot{r}}{\dot{r}}= -(z-1)(2z-3)r^{2(z-2)}M^2
\end{equation}
We get
\begin{eqnarray}
A_{00}&=& -2z(z-1)r^{2(z-2)}M^2\\
A_{ii}&=& -(z-1)r^{2(z-2)}M^2
\end{eqnarray}

From the geodesic equation, one can see that, for $z \geq 1$, near $r\approx 0$, $ r \propto u$ (Here we note that in going from Rosen to Brinkmann coordinates we have $U=u$). So the profile function $A_{ab}(u)$ has the same behaviour. For $z=2$, we get a constant wave profile and for $z >2$, there are no divergences. Note that the null energy condition requires $z \geq 1$ (for positive $z$) and hence $A_{ab}$ are all negative.

\section{ Hyperscaling violating Lifshitz spacetimes} \label{s5}
\setcounter{equation}{0}

In this section, we consider hyperscaling violating Lifshitz geometry and its Penrose limit. Hyperscaling violating Lifshitz spacetime is given by
\begin{equation}
ds^2=\frac{1}{r^{2\theta/d}}\left(-r^{2z}dt^2+\frac{dr^2}{r^2}+r^2d\vec{x}^2\right)
\end{equation}
In appendix B, we calculate Penrose limit for this metric. Here, we will work in a slightly different coordinate system. In \cite{Copsey:2010ya}, possible singularities in hyperscaling violating Lifshitz spacetimes were exhaustively studied and constraints due to null energy condition were analyzed. To match with their conventions, let us perform the following coordinate transformation
\begin{eqnarray}
\theta/d&=&1-1/n\\
z&=&(m+n-1)/n\\
r&=&r'^n\\
t&=&nt'\\
x&=&nx'\\
\end{eqnarray}
we get the metric as
\begin{equation}
ds^2=n^2\left(-r'^{2m}dt'^2+\frac{dr'^2}{r'^{2n}}+r'^2d\vec{x'}^2\right)
\end{equation}

According to the null energy condition
\begin{equation}
R_{\mu\nu}l^\mu l^\nu \geq 0
\end{equation}
(where {\bf l} is a null vector) we have, following \cite{Copsey:2010ya},
\begin{eqnarray}
m&\geq&n\\
(m-1)\left[m+n+d-3\right]&\geq&0
\end{eqnarray}

Case $m=n=1$ just corresponds to $AdS$. The Lagrangian for this case can be written as
\begin{equation}
{\cal L}=\frac{n^2}{2}\left[-r'^{2m}\dot{t}^2+\frac{\dot{r}'^2}{r'^{2n}}+r'^2\dot{x}_1'^2\right]
\end{equation}
Here, derivative is with respect to affine parameter $U$. The geodesic equations are
\begin{eqnarray}
\dot{t}&=&\frac{E}{r'^{2m}}\\
\dot{x}_1'&=&\frac{P}{r'^2}
\end{eqnarray}
Using the null geodesic condition ${\cal L}=0$ we get
\begin{equation}
\dot{r}'^2=r'^{2(n-1)}\left[\frac{E^2}{r'^{2(m-1)}}-P^2\right]
\end{equation}
Integrating the geodesic equations(considering $r'\equiv r'(U)$),we get
\begin{eqnarray}
t&=&\int\frac{E}{r'^{2m}}dU+T\\
x_1'&=&\int\frac{P}{r'^2}dU+X\\
r'&=&\int\dot{r}'dU
\end{eqnarray}
Here $T$ and $X$ are integration constants along the geodesic with parameter $U$. Hence
\begin{eqnarray}
dt&=&\frac{E}{r'^{2m}}dU+dT\\
dx_1'&=&\frac{P}{r'^2}dU+dX\\
dr'&=&r'^{n-1}\sqrt{\frac{E^2}{r'^{2(m-1)}}-P^2}dU
\end{eqnarray}
So in terms of $U,T$ and $X$ we have the metric as
\begin{equation}
ds^2=2dU\left(-EdT+PdX\right)-r'^{2m}dT^2+r^2dX^2+\sum_{i=2}^d r'^2dx_i'^2
\end{equation}
Defining
\begin{equation}
V=-ET+PX
\end{equation}
the metric can be written as
\begin{equation}
2dUdV-\frac{r'^{2m}}{E^2}\left(\frac{E^2}{r'^{2(m-1)}}-P^2\right)dX^2-\frac{r'^{2m}}{E^2}dV^2+\frac{2Pr'^2}{E^2}dVdX+\sum_{i=2}^d r'^2dx_i'^2
\end{equation}
Taking the Penrose limit we get
\begin{equation}
ds_\gamma^2=2dUdV-\frac{r'^{2m}}{E^2}\left(\frac{E^2}{r'^{2(m-1)}}-P^2\right)dX^2+\sum_{i=2}^d r'^2dx_i'^2
\end{equation}
or
\begin{equation}
ds_\gamma^2=2dUdV+\frac{\dot{r}'^2r'^{2(m-n+1)}}{E^2}dX^2+\sum_{i=2}^d r'^2dx_i'^2
\end{equation}
Comparing with the metric of the form
\begin{equation}
ds^2=2dUdV+\sum_{i=1}^d (e_idx_i)^2
\end{equation}
we can read off the vielbeins. The wave profile we get on converting the metric to Brinkmann coordinate system is given by
\begin{equation}
A_{ij}=\delta_{ij}\frac{\ddot{e}_i}{e_i}
\end{equation}
Therefore using
\begin{eqnarray}
\dot{r}'^2&=&E^2r'^{2(n-m)}-P^2r'^{2(n-1)}\\
\ddot{r}'&=&E^2(n-m)r'^{2n-2m-1}-P^2(n-1)r'^{2n-3}\\
\frac{\dddot{r}'}{\dot{r}'}&=&E^2(n-m)(2n-2m-1)r'^{2(n-m-1)}-P^2(n-1)(2n-3)r'^{2(n-2)}
\end{eqnarray}
The wave profile we get is
\begin{equation}
A_{11}=E^2(n-m)r'^{2(n-m-1)}-P^2m(n+m-2)r'^{2(n-2)}
\end{equation}
and for $i>1$ we get
\begin{equation}
A_{ii}=E^2(n-m)r'^{2(n-m-1)}-P^2(n-1)r'^{2(n-2)}
\end{equation}

Near $r \approx 0$, we see that dominant term in $A_{ab}$ is (converting from $r$ to $u$ using radial geodesic equation)
 \begin{equation}
A_{ab}(u)=\delta_{ab}\frac{n-m}{n(m-n+1)}\frac{1}{u^2}
\end{equation}
where $(U=u)$. 

The null energy condition suggests that the Penrose limit wave profile of the hyperscaling violating metric will be negative at least in the near singularity limit. There are subdominant pieces and in the special case, $n=m$ when leading divergent piece is not there, we get weaker singularities
 which go like $A_{ab} \propto u^{2(n-2)}$. For $m=n \geq 2$, we get regular Penrose limit metric. This agrees with the analysis of
 \cite{Copsey:2010ya}. Smooth extensions for this nonsingular hyperscaling violating Lifshitz geometry were obtained in \cite{Lei:2013apa}. Even in
 these nonsingular cases, it is possible that even though the metric is nonsingular, derivatives of the metric may be singular. This will happen when $2(n-2)$ is
 not an integer. Since wave profile $A_{ab}$ of the Penrose limit metric captures the Riemann tensor of original metric, non-integer powers of $u$, can
 lead to divergences in derivatives of Riemann tensor.

 For $m=n$ and $1 \leq n <2$, we have singular Penrose limit metric but the singularity is weaker than $1/u^2$. Such cases are called weak null
 singularities and were discussed in \cite{David:2003vn}. In this case, even though there are divergent tidal forces at $u=0$, string propagation is
 smooth and one can continue the metric beyond the singularity. Physically, the distortion suffered by a freely falling observer is finite and a falling
 string doesn't become infinitely excited while crossing $u=0$. In \cite{Blau:2004yi}, using dominant energy condition, it was conjectured that there
 is a universal $u^{-2}$  dominant near-singularity behaviour. In plane wave spacetimes obtained here, we also got $u^{-2}$ behavior for near-singularity region in Lifshitz case and general $m \neq n$ hyperscaling violating Lifshitz spacetimes. But the class $m=n$ and $1 \leq n <2$ 
seems to lead to weaker singularity for metrics satisfying the null energy condition. We plan to investigate this question later. 
In the appendix, we derive the Penrose limit metric for the hyperscaling violating spacetimes in a different coordinate system. 
\section{ Hyperscaling violating Schrodinger spacetime } \label{s6}
\setcounter{equation}{0}

In this section, we study Penrose limit metric for Schrodinger spacetimes with hyperscaling  violation. In \cite{Ogawa:2011bz,Dong:2012se,Kim:2012nb}, various aspects of these spacetimes have been discussed. 
\begin{equation}
ds^2=r^{-k}\left\{-r^{2z}dt^2+r^2\left(-2dtd\zeta+d\vec{x}^2\right)+\frac{dr^2}{r^2}\right\}
\end{equation}

\subsection{Null energy condition} \label{s6.1}
Null energy condition can be used to constrain the parameters of this metric to physically acceptable values.  Even though this has been discussed in
\cite{Ogawa:2011bz,Dong:2012se,Kim:2012nb}, we present this in our conventions. For the null vector $l^a$  the condition becomes
\begin{equation}
R_{ab}l^al^b\geq0
\end{equation}
For the $(d+3)$ dimensional spacetime, non zero components of the Ricci tensor are
\begin{eqnarray}
R_{tt}&=&\frac{1}{4}\left[8(z^2+1)+4(d-2)z-2(d+2)k+(d+1)k(k-2z)\right]r^{2z}\\
R_{rr}&=&(d+2)\frac{k-2}{2r^2}\\
R_{xx}&=&-\frac{1}{4}\left[(d+1)k^2-2(2d+3)k+4(d+2)\right]r^2\\
R_{t\zeta}=R_{\zeta t}&=&\frac{1}{4}\left[(d+1)k^2-2(2d+3)k+4(d+2)\right]r^2
\end{eqnarray}
So the null energy condition becomes
\begin{equation}
\begin{split}
\frac{1}{4}\left[8(z^2+1)+4(d-2)z-2(d+2)k+(d+1)k(k-2z)\right]r^{2z}(l^t)^2+(d+2)\frac{k-2}{2r^2}(l^r)^2 \\
+\frac{1}{2}\left[(d+1)k^2-2(2d+3)k+4(d+2)\right]r^2 l^tl^\zeta-\frac{1}{4}\left[(d+1)k^2-2(2d+3)k+4(d+2)\right]r^2|l_{tr}|^2 
\end{split}
\end{equation}
This can be written as
\begin{equation}
\begin{split}
-\frac{1}{4}\left[(d+1)k^2-2(2d+3)k+4(d+2)\right]\left[-r^{2z}(l^t)^2+r^2\left(-2l^tl^\zeta+|l_{tr}|^2\right)+\frac{(l^r)^2}{r^2}\right] \\
+\frac{1}{4}\left[8z^2+4(d-2)z-2k(d+1)(z-1)-4d\right]r^{2z}(l^t)^2+\frac{1}{4}(d+1)(k^2-2k)r^2(l^r)^2 
\end{split}
\end{equation}
Since the vector {\bf l} is null, we have the null energy condition as
\begin{eqnarray}
(z-1)\left[8z+4d-2k(d+1)\right] &\geq& 0\\
k(k-2) &\geq& 0
\end{eqnarray}
where for $z\geq1$ we get
\begin{equation}
k\leq\frac{2(2z+d)}{d+1}
\end{equation}

\subsection{Penrose Limit} \label{s6.2}
To obtain the Penrose limit, we find geodesics using the Lagrangian  given by
\begin{equation}
{\cal L}=\frac{1}{2}\left(-r^{-k+2z}\dot{t}^2-2r^{-k+2}\dot{t}\dot{\zeta}+ r^{-k+2}\dot{x}_1^2+r^{-(k+2)}\dot{r}^2\right)
\end{equation}
Equations of motion are
\begin{eqnarray}
\dot{t}&=&Mr^{(k-2)}\\
\dot{\zeta}&=&Er^{(k-2)}-Mr^{2z+k-4}\\
\dot{x}_1&=&Pr^{(k-2)}
\end{eqnarray}
Using the null geodesic condition i.e. ${\cal L}=0$ we get
\begin{equation}
\dot{r}^2=(2ME-P^2)r^{2k}-M^2r^{2(z+k-1)} \label{radialgeo}
\end{equation}
On integrating along the given null geodesic we get
\begin{eqnarray}
t&=&M\int r^{(k-2)}dU+T \ \ , \ \ \zeta=\int\left\{Er^{(k-2)}- Mr^{2z+k-4}\right\}dU+\lambda \\
x_1&=&P\int r^{(k-2)}dU+X \ \ , \ \ r=\int\dot{r}dU
\end{eqnarray}
Here $T,\lambda$ and $X$ are integration constant along the geodesic with affine parameter $U$. Hence
\begin{eqnarray}
dt&=&Mr^{(k-2)}dU+dT\\
d\zeta&=&\left\{Er^{(k-2)}- Mr^{2z+k-4}\right\}dU+d\lambda\\
dx_1&=&Pr^{(k-2)}dU+dX\\
dr&=&\dot{r}dU
\end{eqnarray}
Rewriting the metric in terms of $T,\lambda$ and $X$ we get
\begin{equation}
ds^2=2dU(-EdT-Md\lambda+PdX)-r^{2z-k}dT^2+r^{-k+2}dX^2-2r^{-k+2}dTd\lambda+\sum_{i=2}^d r^{-k+2}dx_i^2
\end{equation}
We define 
\begin{equation}
V=-ET-M\lambda+PX
\end{equation}
Trading $\lambda$ for $V$ and taking Penrose limit we get
\begin{equation}
ds^2=2dUdV+\left(2r^{-k+2}\frac{E}{M}-r^{2z-k}\right)dT^2-2\frac{P}{M}r^{-k+2}dTdX+r^{-k+2}dX^2+r^{-k+2}\sum_{i=2}^d dx_i^2
\end{equation}
Let us define
\begin{equation}
X_1=X-\frac{P}{M}T
\end{equation}
On trading $X$ for $X_1$ in the above equation we get
\begin{equation}
ds^2=2dUdV+\left(\frac{2ME-P^2}{M^2}r^{-k+2}-r^{2z-k}\right)dT^2+r^{-k+2}dX_1^2+r^{-k+2}\sum_{i=2}^d dx_i^2
\end{equation}

After renaming rest of $x_i$ as $X_i$ and using geodesic equation, we get 
\begin{equation}
ds^2=2dUdV+\frac{r^{-3k+2}\dot{r}^2}{M^2}dT^2+r^{-k+2}\sum_{i=1}^d dX_i^2
\end{equation}
For the metric of this type, the wave profile we get on converting it to Brinkmann coordinate system is
\begin{equation}
A_{ij}=\delta_{ij}\frac{\ddot{e}_i}{e_i}
\end{equation}
In this case we have
\begin{equation}
\frac{\ddot{e}_0}{e_0}=\frac{\dddot{r}}{\dot{r}}+3\left(-\frac{3}{2}k+1\right)\frac{\ddot{r}}{r}+\frac{3}{2}k\left(\frac{3}{2}k-1\right)\frac{\dot{r}^2}{r^2}
\end{equation}
So
\begin{equation}
A_{00}=\frac{k}{2}\left(\frac{k}{2}-1\right)(2ME-P^2)r^{2(k-1)}-\left\{2z\left(z-1-\frac{k}{4}\right)+k\left(1-\frac{k}{4}\right)\right\}M^2r^{2(z+k-2)}
\end{equation}
and for $i\neq0$ we have
\begin{equation}
\frac{\ddot{e}_i}{e_i}=\frac{k}{2}\left(\frac{k}{2}-1\right)\frac{\dot{r}^2}{r^2}+\left(1-\frac{k}{2}\right)\frac{\ddot{r}}{r}
\end{equation}
So
\begin{equation}
A_{ii}=\frac{k}{2}\left(1-\frac{k}{2}\right)(2ME-P^2)r^{2(k-1)}-\left\{z\left(1-\frac{k}{2}\right)-\left(1-\frac{k}{2}\right)^2\right\}M^2r^{2(z+k-2)}
\end{equation}

For $k=0$ i.e. Schrodinger spacetime without hyperscaling violation, nonsingular metric for $z \geq 2$ and metric extension was done in \cite{Blau:2009gd}. We obtained the Penrose limits for these cases earlier. With hyperscaling violation i.e non-zero $k$,for the wave profile to be finite at $r=0$ (i.e. Penrose limit metric to be non-singular) we need to have
\begin{eqnarray}
k-1 &\geq& 0\\
k+z-2 &\geq& 0
\end{eqnarray}
According to the null energy condition either $k\leq0$ or $k\geq2$. So for the plane wave metric to  be regular at origin we need to have $k\geq2$.
\vskip 0.1in
For $z\geq1$ and $k>2$, in the near $r=0$ limit we have
\begin{eqnarray}
A_{00}\approx  \frac{k}{2}\left(\frac{k}{2}-1\right)(2ME-P^2)r^{2(k-1)}\\
A_{ii}\approx \frac{k}{2}\left(1-\frac{k}{2}\right)(2ME-P^2)r^{2(k-1)}
\end{eqnarray}
Using
\begin{equation}
\dot{r}^2\approx (2ME-P^2)r^{2k}
\end{equation}
i.e.
\begin{equation}
r^{k-1}=-\frac{1}{(k-1)(2ME-P^2)^{1/2}U}
\end{equation}
The wave profile becomes
\begin{eqnarray}
A_{00}=\frac{k(k-2)}{4(k-1)^2}u^{-2}\\
A_{ii}=\frac{k(2-k)}{4(k-1)^2}u^{-2}
\end{eqnarray}

where $(U=u)$.

Since $r$ and $u$ are inversely related, $r \rightarrow 0$ corresponds to $u \rightarrow \infty$ and hence the wave profile is not singular as $r \rightarrow 0$. Since $A_{ab}$ has one negative eigenvalue, there might be stability issues. 
For $k=2$ and $z>1$, we see that the Penrose limit metric is quite simple, with only non-zero component of $A_{ab}$ being $A_{00} = -(z-1)(2z-1)M^2 r^{2z}$.
\subsection{k=2 Hyperscaling violating Schrodinger Geometry} \label{s6.3}

For $k=2$, hyperscaling violating Schrodinger metric is 

\begin{equation}
ds^2=-r^{2z-2}dt^2+ \left(-2dtd\zeta+d\vec{x}^2\right)+\frac{dr^2}{r^4}
\end{equation}

For $z=1$, we just get part of flat space which can be extended to whole of flat space. This is expected since $z=1$ Schrodinger metric corresponds to $AdS$ which is conformal to flat space. Hyperscaling violating factor just cancels this conformal factor. 
For $z=2$, we can extend this metric beyond $r=0$  by following coordinate transformations, which are simplified forms of transformations used to go to Penrose adapted coordinates 
\begin{eqnarray}
t&=&MU+T\\
\zeta&=& \int (E - Mr^2)dU+\lambda\\
r&=&\ \int \dot{r} dU
\end{eqnarray}
where dot means derivative with respect to $U$ and we use \ref{radialgeo} with $P=0$ to convert from $r$ to $U$. In terms of these coordinates $r \rightarrow 0$ corresponds to $U \rightarrow -\infty$. Metric becomes
\begin{equation}
ds^2=- r(U)^{2}dT^2+ \left(-2EdTdU+d\vec{x}^2\right) -2d\lambda (MdU +dT) 
\end{equation}

In these coordinates, metric can be extended beyond $r =0$ as metric components remain smooth there and determinant of the metric is non-zero as $r \rightarrow 0$.  

\section{Conclusions} \label{s7}

We have obtained plane wave metrics corresponding to several geometries which have been used recently in giving dual descriptions of many non-relativistic field theories. It is interesting that Penrose limit metrics are well defined even though the dual field theory is non-relativistic. One feature of the Penrose limit metrics of these geometries is that singularities in the original metrics become manifest in the plane wave limit. This is to be expected since pp-curvature singularities are given by divergences in components of Riemann tensor in parallely
 propagated frame and the wave profile in Brinkmann coordinates is directly related to Riemann tensor in parallely propagated frame . We have not
 considered dual field theory of these Penrose limit metrics in this paper. But based on our experience with $AdS$, we would expect the Penrose limit to 
describe a certain sector of states in field theory. It would be interesting to flesh out the details of the analogue of BMN limit for these cases. 
Since plane wave spacetimes are quite tractable for detailed string theory and field theory considerations, one can do a lot more in this limit than 
in general non-relativistic metrics. We are currently working on various aspects of this correspondence, including string theory in these plane wave backgrounds and will report on them in the near future. 

\appendix

\sectiono{Review of Penrose Limit}
\label{sec:appendix A}

In this section, we review methods to obtain Penrose limits, especially the covariant characterization of Penrose limit as discussed in detail in 
\cite{Blau:2003dz}and \cite{Blaureview}. This plane wave depends upon the metric we start with and the choice of the null geodesic along which the observer is moving.
\subsection{Penrose limit via adapted coordinates} 

To find the Penrose limit of a given metric for a given geodesic congruence we need to write the metric given to us in the Penrose Adapted Coordinates. 
\begin{equation}
ds_\gamma^2=2dUdV+a(U,V,Y^k)dV^2+2b_i(U,V,Y^k)dVdY^k+g_{ij}(U,V,Y^k)dY^idY^j
\end{equation}
where $U$ plays the role of the affine parameter along the chosen null geodesic.  Then we have to perform a coordinate transformation which is same as scaling the metric followed by a boost.
\begin{equation}
(U,V,Y^k)\rightarrow(u,\lambda^2v,\lambda y^k)\;\;\;\lambda{\cal 2}R
\end{equation}
The Penrose limit metric is defined as
\begin{equation}
ds^2=\lim_{\lambda\rightarrow0}\lambda^{-2}ds^2_{\gamma\lambda}
\end{equation}
where $ds^2_{\gamma\lambda}$ is the metric we get upon performing the above defined coordinate transformation. So after taking the Penrose limit we get the metric
\begin{equation}
ds^2=2dudv+g_{ij}(u)dy^idy^j
\end{equation}
This is the familiar plane wave metric in Rosen coordinates. For most purposes, a second coordinate system, called Brinkmann coordinates, is better suited to represent plane waves. In this coordinate system
\begin{equation}
ds^2=2dudv+A_{ab}x^ax^bdu^2+d\vec{x}^2
\end{equation}

In Brinkmann coordinates all the information of the wave is contained in the coefficient of $du^2$ which is $A_{ab}$. One more thing we need is the coordinate transformation from Rosen form to Brinkmann form. We see that the transverse part of the metric is flat in Brinkmann coordinate and is not in general flat in Rosen coordinate so we can change the variable as
\begin{equation}
x^a=E^a_iy^i
\end{equation}
where $E^a_i$ are vielbein for $g_{ij}$ i.e.
\begin{equation}
g_{ij}=\delta_{ab}E^a_iE^b_j
\end{equation}
So
\begin{eqnarray}
g_{ij}dy^idy^j&=&(dx^a-\dot{E}^a_iE^i_cx^cdU)(dx^b-\dot{E}^bjE^j_dx^ddU)\delta_{ab}\\
&=&d\vec{x}^2-2\dot{E}_{ai}E^i_cx^cdUdx^a+ \delta_{ab}\dot{E}^a_i\dot{E}^b_jE^i_cE^j_dx^cx^ddU^2
\end{eqnarray}
Choosing the vielbeins such that
\begin{equation}
\dot{E}_{ai}E^i_b=\dot{E}_{bi}E^i_a \label{viel cond}
\end{equation}
and
\begin{equation}
V=v+\frac{1}{2}\dot{E}_{ai}E^i_bx^ax^b
\end{equation}
For $u=U$, we can write $dUdV$ as
\begin{eqnarray}
2dUdV&=&2dUd(v+\frac{1}{2}\dot{E}_{ai}E^i_bx^ax^b)\\
&=&2dudv+\frac{d}{du}\left(\dot{E}_{ai}E^i_b\right)x^ax^bdu^2+2\dot{E}_{ai}E^i_bx^bdudx^a
\end{eqnarray}
Therefore total metric $2dUdV+g_{ij}dy^idy^j$ can be written as
\begin{equation}
ds^2=2dudv+\left[\frac{d}{du}\left(\dot{E}_{ai}E^i_b\right)+ \delta_{cd}\dot{E}^c_i\dot{E}^d_jE^i_aE^j_b\right]x^ax^bdu^2+d\vec{x}^2
\end{equation}
So we have got the structure of the plane wave in Brinkmann Coordinates, now we need to simplify the coefficient of $du^2$ to get the wave profile $A_{ab}$
$$
\frac{d}{du}\left(\dot{E}_{ai}E^i_b\right)+\delta_{cd}\dot{E}^c_i\dot{E}^d_jE^i_aE^j_b = \ddot{E}_{ai}E^i_b+\dot{E}_{ai}\dot{E}^i_b+ \delta_{cd}\dot{E}^c_i\dot{E}^d_jE^i_aE^j_b
$$
where the last two term in the RHS of the above equation
$$
\dot{E}_{ai}\dot{E}^i_b+\delta_{cd}\dot{E}^c_i\dot{E}^d_jE^i_aE^j_b = \dot{E}_{ai}\dot{E}^i_b+ \delta_{cd}E^c_iE^d_j\dot{E}^i_a\dot{E}^j_b = 
\dot{E}^i_b(\dot{E}_{ai}+\delta_{cd}E^c_jE^d_i\dot{E}^j_a)
$$
where it can be shown that the term in the bracket is zero, this can be done by multiplying the term with $E^i_b$
$$
\dot{E}_{ai}E^i_b+\delta_{cd}E^c_jE^d_i\dot{E}^j_aE^i_b = \dot{E}_{ai}E^i_b+E_{bi}\dot{E}^i_a
$$
Using equation \ref{viel cond} we get the above as
$$
 \dot{E}_{ai}E^i_b+E_{ai}\dot{E}^i_b=\frac{d}{du}\left(\delta_{ab}\right)=0
$$
and therefore
\begin{equation}
A_{ab}(u)=\ddot{E}_{ai}E^i_b
\end{equation}
and hence the metric in Brinkmann coordinates takes the form
\begin{equation}
2dudv+\ddot{E}_{ai}E^i_bx^ax^bdu^2+d\vec{x}^2
\end{equation}
 
In this next subsection, we will review the covariant method which gives metric in Brinkmann form directly. Once we have the given metric in Penrose adapted coordinates, we can see that the effect of taking Penrose limit is same as neglecting the $dV^2$ and $dVdY^k$ term from the metric and keeping the remaining term as it is. Also $g_{ij}(U)$ is the restriction of the $g_{ij}(U,V,Y^k)$ on the null geodesic $\gamma$. So on taking the Penrose limit we are actually observing an infinitesimal region of the spacetime which is near the null geodesic $\gamma$ and expanding it to form the entire spacetime.

\subsection{Covariant description of the Penrose limit} \label{s2.2}
On taking the Penrose limit for a given metric ($g_{\mu\nu}$) and null geodesic ($\gamma$) congruence, what we are interested in is the wave profile ($A_{ab}$) of the limiting plane wave metric. In this section, we will see the covariant way of calculating the wave profile instead of going through the several coordinate transformation and rescaling of the metric. This method was presented in \cite{Blau:2003dz} and we refer to that and \cite{Blaureview} for details. \\
One can see that the profile $A_{ab}=-R_{uaub}$ in Brinkmann coordinates. In terms of the plane wave in Rosen coordinates, it is
\begin{equation}
A_{ab}=-E^i_aE^j_bR_{iUjU}
\end{equation}
where $E^a_i$ are the vielbeins for the transverse metric in the Rosen coordinate ($g_{ij}$) and satisfies the symmetry condition
\begin{equation}
\dot{E_{ai}}E^i_b=\dot{E_{bi}}E^i_a
\end{equation}
For the metric in Penrose adapted coordinate, the component of the Riemann tensor $R^i_{UjU}$ is given as
\begin{equation}
R^i_{UjU}=-\left(\partial_U\Gamma^i_{jU}+\Gamma^i_{kU}\Gamma^k_{jU}\right)
\end{equation}
This is independent of $a$ and $b_i$ and depends only upon $g_{ij}$ and its $U$ derivative. (This can easily be seen by just expanding each of the Christoffel symbol in the above expression.). Hence on taking the Penrose limit, we will have
\begin{equation}
\bar{R}^i_{UjU}=R^i_{UjU}|_\gamma
\end{equation}
In order to look for a covariant description, we would like to write the metric given to us as
\begin{equation}
ds^2=2E^+E^-+\delta_{ab}E^aE^b
\end{equation}
where $E^A$ are parallel along the null geodesic congruence i.e.$\nabla_UE^A_\mu=0$. We choose one of the leg to be the tangent along the null goedesic
$E^+=\partial_U $.

Finally we get our wave profile of the Penrose limit metric as $A_{ab}=\frac{d}{du}B_{ab}+B_{ac}B^c_b$
where $B_{ab}=E_a^\mu E_b^\nu \nabla_\mu\partial_\nu S $. Although we have derived these results in Penrose adapted coordinates, but what we have got at the end has only vielbeins indices and all the coordinate indices has been summed up. So this quantity is a scalar. Hence we can use any coordinate system to compute these quantities and we will get the same result at the end.\\

\sectiono{Hyperscaling Violating Lifshitz again}
\label{sec:appendix B}
In this section, we consider consider Penrose limit of hyperscaling
 violating Lifshitz metric in a slightly different coordinate system. Unlike section $5$, we use covariant method for Penrose limit here. 

\begin{equation}
ds^2=\frac{1}{r^{2\theta/d}}\left(-r^{2z}dt^2+\frac{dr^2}{r^2}+r^2d\vec{x}^2\right)
\end{equation}
Lagrangian for this system is
\begin{equation}
{\cal L}=\frac{1}{2r^{2\theta/d}}\left(-r^{2z}\dot{t}^2+\frac{\dot{r}^2}{r^2}+r^2\dot{x}_1^2\right)
\end{equation}
Since the slice of this spacetime at constant $r$ and $t$ is just a flat space, we can choose our coordinate system in such a way that only one of the $x$ vary along the geodesic. Solving the Euler Lagrange equation for $t$, we get
\begin{equation}
\dot{t}=\frac{E}{r^{2(z-\theta/d)}}
\end{equation}
where $-E$ is the conserved quantity corresponding to the $t$ coordinate. Solving the Euler Lagrange equation for $x_1$ we get
\begin{equation}
\dot{x}_1=\frac{P}{r^{2(1-\theta/d)}}
\end{equation}
where $P$ is the momentum corresponding to the $x_1$ coordinate. For the equation in $r$, we can use the condition for the null geodesic, i.e. ${\cal L}=0$ which gives
\begin{equation}
\dot{r}^2=r^{4\theta/d}\left[\frac{E^2}{r^{2(z-1)}}-P^2\right]
\end{equation}
Hamilton Jacobi function for the above case is
\begin{equation}
S=-Et+Px_1+\rho(r)
\end{equation}
where
\begin{equation}
\rho'(r)=\frac{\partial{\cal L}}{\partial\dot{r}}=\frac{\dot{r}}{r^{2(1+\theta/d)}}
\end{equation}
Now we need to write the above metric in parallely propagated frame basis
\begin{equation}
ds^2=2E^+E^-+\delta_{ab}E^aE^b
\end{equation}
where the leg $E_+$ is directed along the tangent to the given null geodesic.
\begin{equation}
E_+=\dot{t}\partial_t+\dot{x_1}\partial_{x_1}+\dot{r}\partial_r
\end{equation}
For $i=2,3,\dots,n$ we can choose
\begin{equation}
E^i=r^{(1-\theta/d)}dx_i
\end{equation}
\begin{eqnarray}
B_{ij}&=&E_i^{x_i}E_j^{x_j}\nabla_{x_i}\partial_{x_j}S\\
&=&-E_i^{x_i}E_j^{x_j}\Gamma^\alpha_{x_ix_j}\partial_\alpha S
\end{eqnarray}
where $\alpha$ can only be $t$, $r$, and $x_1$ or else $\partial_\alpha S=0$. Putting values in this, we get
\begin{equation}
B_{ij}=\delta_{ij}(1-\theta/d)\frac{\dot{r}}{r}=\delta_{ij}\partial_u\log(r^{1-\theta/d})
\end{equation}
To know $B_{11}$ we need to know the trace of B for which, we have the formula
\begin{equation}
Tr(B)=\frac{1}{\sqrt{-g}}\partial_\mu(\dot{x}^\mu\sqrt{-g})
\end{equation}
where $\sqrt{-g}=r^{D(1-\theta/d)+z-3}$ where $D$ is the dimension of the spacetime given above. Therefore
\begin{equation}
Tr(B)=\partial_u\log(\dot{r}r^{D(1-\theta/d)+z-3})
\end{equation}
So $B_{11}=Tr(B)-(B_{22}+B_{33}+\dots+B_{nn})$
\begin{eqnarray}
B_{11}&=&\partial_u\log\left[\frac{\dot{r}r^{D(1-\theta/d)+z-3}}{r^{(D-3)(1-\theta/d)}}\right]\\
&=&\partial_u\log(\dot{r}r^{z-3\theta/d})
\end{eqnarray}
$B_{1i}$ for any $i$ between 2 and $n$ will be 0 as can be easily checked. From these, we get the profile function as
\begin{eqnarray}
A_{11}&=&\frac{1}{\dot{r}r^{z-3\theta/d}}\partial_u^2\left(\dot{r}r^{z-3\theta/d}\right) \nonumber \\
&=&E^2\left(1-\frac{\theta}{d}\right)\left(1-z+\frac{\theta}{d}\right)r^{2(2\theta/d-z)}+P^2\left[z -z^2 -\frac{\theta}{d} +\frac{\theta^2}{d^2}\right]r^{2(2\theta/d-1)}
\end{eqnarray}
and for $i,j=2,3,\dots,n$
\begin{eqnarray}
A_{ij}&=&\delta_{ij}\frac{1}{r^{1-\theta/d}}\partial_u^2\left(r^{1-\theta/d}\right) \nonumber\\
&=&E^2\left(1-\frac{\theta}{d}\right)\left(1-z+\frac{\theta}{d}\right)r^{2(2\theta/d-z)}-P^2\frac{\theta}{d}\left(1-\frac{\theta}{d}\right)r^{2(2\theta/d-1)}
\end{eqnarray}

\end{document}